\documentclass[twocolumn,showpacs,preprintnumbers,amsmath,amssymb]{revtex4}

\usepackage{graphicx}
\usepackage{dcolumn}
\usepackage{bm}

\newcommand\ignore[1]{}
\def\one{{\,\hbox{1\kern-.8mm l}}}

\def\Tr{{\rm Tr\, }}

\newcommand{\Cset}{{\,\,{{{^{_{\pmb{\mid}}}}\kern-.45em{\mathrm C}}}}}

\newcommand{\be}{\begin{equation}}
\newcommand{\ee}{\end{equation}}
\newcommand{\bea}{\begin{eqnarray}}
\newcommand{\eea}{\end{eqnarray}}

\def\a{\alpha}

\def\d{\partial}

\begin{document}



\title{A Nonabelian Particle-Vortex Duality}

\preprint{QGaSLAB-15-02}
\author{Jeff Murugan$^{1}$}\email{jeff.murugan@uct.ac.za}
\author{Horatiu Nastase$^{2}$}\email{nastase@ift.unesp.br}
\affiliation{${}^{1}$The Laboratory for Quantum Gravity \& Strings, Department of Mathematics and Applied Mathematics, University of Cape Town,Private Bag, Rondebosch 7700, South Africa}
\affiliation{${}^{2}$Instituto de F\'{i}sica Te\'{o}rica, UNESP-Universidade Estadual Paulista, Rua Dr. Bento T. Ferraz 271, Bl. II, S\~ao Paulo 01140-070, SP, Brazil}
\date{\today}

\begin{abstract}
We define a nonabelian particle-vortex duality as a $3-$dimensional analogue of the usual $2-$dimensional worldsheet nonabelian T-duality. 
The transformation is defined in the presence of a global $SU(2)$ symmetry and, although derived from a string theoretic setting, we formulate it  
generally. 
We then apply it to so-called ``semilocal strings" in an 
$SU(2)_{G}\times U(1)_{L}$ gauge theory, originally discovered in the context of cosmic string physics.

\end{abstract}


\maketitle

\section{Introduction}
\noindent
Beginning with the remarkable correspondence between the sine-Gordon and massive Thirring models \cite{Coleman:1974bu}, dualities have 
played a crucial role in the modern understanding of quantum field theories. Indeed, they have been an indispensible tool in the understanding 
of both strongly coupled systems as well as various nonperturbative problems. This was certainly the case, for instance, for Seiberg and Witten's 
landmark study of (3+1)-dimensional, ${\cal N}=2$ supersymmetric gauge theory \cite{Seiberg:1994rs,Seiberg:1994aj}, 
where electric-magnetic duality (a generalized form of the usual electric-magnetic duality of Maxwell electrodynamics) that exchanges particles 
with monopoles, was essential in fully solving the low energy theory. In that $(3+1)-$dimensional case, even though an explicit path integral 
transformation exists only for the abelian case, the duality is understood as being essentially non-abelian in the sense of acting on the full non-abelian theory. 

One duality which has received considerably less attention occurs in $(2+1)-$dimensional gauge theories and exchanges particles with topological 
solitons, specifically vortices \cite{Lee:1991jt}. One possible reason for the dearth of literature on the subject could be that its utility lies primarily in 
condensed matter systems which, being usually non-relativistic are much less susceptible to the powerful relativistic
methods employed in high energy theory. Another is likely the fact that the duality was generally less well-defined than its $(3+1)-$ dimensional 
counterpart. To the best of our knowledge, {\it particle-vortex duality} has, until now, only been defined in the context of abelian gauge theories, 
exhibiting Neilsen-Olesen-like vortices. In \cite{Murugan:2014sfa}, this duality was 
defined as a path integral transformation in a manifestly symmetric way, and embedded into a planar $\mathcal{N}=6$ Chern-Simons-matter 
theory commonly known as the ABJM model, which is itself known to be dual to the type IIA superstring 
on an $AdS_4\times\mathbb{CP}^3$ background \cite{Aharony:2008ug}. In this context, the particle-vortex duality of the boundary field theory 
was shown to correspond to an electric-magnetic duality in the bulk. As a final point in \cite{Murugan:2014sfa}, it was speculated that, based 
on the structure of the embedding into the ABJM model, it should be possible to define a nonabelian version that would act on the whole 
non-abelian ABJM model.

In this letter, we show that it is indeed the case that we can define a version of particle-vortex duality that acts on a non-abelian theory, at 
least in a certain restricted sense. Key to our argument are the recent advances in the study of $2-$dimensional {\it non-abelian T-duality} 
acting on the string worldsheet in string theory \cite{de la Ossa:1992vc} (see also \cite{Sfetsos:2010uq,Itsios:2013wd,Kelekci:2014ima} for 
the action of the nonabelian T-duality in supergravity). By generalizing the procedure to $(2+1)-$dimensions, we obtain a non-abelian version 
of particle-vortex duality that acts on gauge theories with a global $SU(2)$, as well as a local symmetry. 
Recognizing that this is precisely the set-up for 
the ``semi-local" vortices found in \cite{Vachaspati:1991dz} (see also \cite{Hindmarsh:1991jq,Gibbons:1992gt}) in the context of cosmic 
strings in the case of a local $U(1)$ symmetry, we explicitly exhibit the action of the nonabelian particle-vortex transformation on these solutions. 

The letter is organized as follows. In section 2 we revisit non-abelian T-duality and its relation to the abelian T-duality, extending it in 
section 3 to three spacetime dimensions, consequently defining a non-abelian particle-vortex duality on a general theory which we illustrate 
with a simple example of a semilocal vortex in section 4. 
This article should be viewed as a proof-of-principle of a phenomenon with potential application from condensed matter to cosmology, with a longer 
companion paper to follow in which we will elaborate further on the duality and provide more substantial examples \cite{MNsoon}.

\section{Nonabelian T-duality}

In string theory, abelian T-duality is a symmetry that acts on a compact dimension as an inversion of its radius, $R\rightarrow \a'/R$. First noted 
at the level of the string spectrum, it was proven to be a symmetry of the perturbative string path integral in \cite{Buscher1988}, where it was 
defined as a duality transformation of the worldsheet action. Specifically, one writes a constrained first order form for the worldsheet action 
for the compact direction, with a Lagrange multiplier implementing the constraint that mixed second derivatives of the compact coordinate 
commute. Then, if instead of eliminating the Lagrange multiplier the original coordinate is integrated out, one obtains a T-dual theory in which 
the Lagrange multiplier plays the role of a new coordinate. This formulation is very similar in spirit to the abelian particle-vortex duality 
transformation at the level of the path integral \cite{Murugan:2014sfa}. 

Initially carried out with commuting abelian isometries, a natural next step was to ``nonabelianize" the transformation. This was first accomplished 
in \cite{de la Ossa:1992vc} with the transformation acting on three coordinates transforming under a (global) $SU(2)$ symmetry, obtaining what 
became known as {\it non-abelian T-duality}. In this section we review the procedure.

Consider the string background with metric and B-field
\bea
ds^2&=&G_{\mu\nu}dx^\mu dx^\nu +2G_{\mu i}dx^\mu L^i+g_{ij}L^iL^j\cr
B&=& B_{\mu\nu}dx^\mu \wedge dx^\nu +B_{\mu i}dx^\mu \wedge L^i+\frac{1}{2}b_{ij}L^{i}\wedge L^j,
\eea
and constant dilaton $\phi=\phi_0$. Here,
\bea
L_1&=& \frac{1}{\sqrt{2}}(-\sin\psi d\theta +\cos\psi \sin \theta  d\phi),\cr
L_2&=&\frac{1}{\sqrt{2}}(\cos\psi d\theta +\sin\psi\sin\theta d\phi),\\
L_3&=&\frac{1}{\sqrt{2}}(d\psi+\cos\theta d\phi)\;,\nonumber
\eea
are $SU(2)$ left-invariant $1-$forms for the Euler angles $(\theta, \phi,\psi)$, such that 
$
dL^i=\frac{1}{2}{f^i}_{jk}L^j\wedge L^k.
$
The angles have the range $0\leq \theta\leq \pi$, $0\leq \phi\leq 2\pi$, $0\leq \psi\leq 4\pi$, and the $SU(2)$ transformations 
act as 
\bea
\delta \theta&=&\epsilon_1\sin\phi+\epsilon_2\cos\phi,\cr
\delta\phi&=&\cot\theta(\epsilon_1\cos\phi-\epsilon_2\sin\theta)+\epsilon_3,\\
\delta\psi&=&\frac{1}{\sin\theta}(-\epsilon_1\cos\phi+\epsilon_2\sin\phi).\nonumber
\eea
Using the normalized Pauli matrices $t^i=\tau^i/\sqrt{2}$, that satisfy $\Tr(t^it^j)=\delta^{ij}$, and the group element
$
g=e^{\frac{i\phi \tau_3}{2}}e^{\frac{i\theta\tau_2}{2}}e^{\frac{i\psi\tau_3}{2}},
$
understood here as a field $g(\tau,\sigma)$ on the string worldsheet, the $1-$forms can be rewritten more conveniently as
$
L^i_\pm=-i\Tr(t^ig^{-1}\d_\pm g).
$
Note that while $g$ is complex, the $L_{i}$ are all real. Then, with
\bea
Q_{\mu\nu}&=&G_{\mu\nu}+B_{\mu\nu},\;\;\;
Q_{\mu i}=G_{\mu i}+B_{\mu i}\cr
Q_{i\mu}&=& G_{i\mu}+B_{i\mu},\;\;\;
E_{ij}=g_{ij}+b_{ij}\;,
\eea
the string worldsheet action in this background takes the globally $SU(2)-$invariant form
\bea
S&=&\int d^2\sigma \left[Q_{\mu\nu}\d_+X^\mu \d_-X^\nu +Q_{\mu i}\d_+X^\mu L_-^i\right.\cr
&&\left.+Q_{i\mu} L_+^i \d_-X^\nu +E_{ij}L_+^i L_-^j\right]\;.
\eea
One can make this invariance local by introducing an $SU(2)$ gauge field $A$ and replacing derivatives with covariant derivatives,
$
\d_\pm g\rightarrow D_\pm g=\d_\pm g-A_\pm g\;,
$
which, in turn, replaces $L_\pm^i$ with 
$
\tilde L_\pm ^i=-i\Tr[t^i g^{-1}D_\pm g].
$
Since we don't want to add a new degree of freedom (the gauge field $A$), we need to impose its triviality as a constraint. 
A good way of doing that is by requiring the field strength to vanish  and enforcing this in the action through a Lagrange multiplier term
$
-i\Tr[vF_{+-}]=-i\epsilon^{\mu\nu}\Tr[vF_{\mu\nu}]\;,
$
where $v=v_i$ is an $SU(2)$ adjoint (a triplet) and the field strength
$
F_{+-}=\d_+A_--\d_-A_+-[A_+,A_-].
$
In this way we obtain a first order action that acts as a {\it master action} for the T-duality.
Integrating out the Lagrange multiplier $v$ leads to $F_{+-}=0$ which, in the absence of any topological issues, leads to a trivial $A$, 
equivalent to  $A=0$, recovering the original theory.

If instead, we integrate out the gauge field $A$ and gauge fix the $SU(2)$ symmetry, we get $A_\pm $ in terms of $v$, and on substituting 
into the master action, obtain the T-dual action. Explicitly, we first partially integrate the Lagrange multiplier term to 
\bea
-i\int \Tr[vF_{+-}]&=&\int \left\{ \Tr[+i(\d_+v)A_--i(\d_-v)A_+]\right.\cr
&&\left.-A_+fA_-\right\}\;,
\eea
where 
$A_+fA_-\equiv A_+^if_{ij}A_-^j$ and $f_{ij}\equiv {f_{ij}}^kv_k.$
Then, gauge fixing the $SU(2)$ to $g=1$, replaces
$
L^i_\pm
$ 
by
$
i\Tr[t^iA_\pm]=iA^i_\pm\;,
$
in the master action, giving
\bea
S&=&\int d^2\sigma\left[Q_{\mu\nu}\d_+ X^\mu \d_- X^\nu+Q_{\mu i}\d_+ X^\mu(+iA_-^i)\right.\cr
&&\left.+Q_{i\mu}\d_-X^\mu (+iA_+^i)+E_{ij}(iA_+^i)(iA_-^j)\right.\cr
&&\left.+i\d_+v_i A_-^i-i\d_-v_i A_+^i-A_+^if_{ij}A_-^j\right].
\eea
After varying this with respect to $A_+$ and $A_-$ and solving the resulting equations of motion, we obtain
\bea
A_-^i&=&-iM_{ij}^{-1}(\d_- v_j-Q_{j\mu}\d_- X^\mu)\cr
A_+^i&=&+iM_{ji}^{-1}(\d_+ v_j+Q_{\mu j}\d_+ X^\mu)\;,
\eea
where $M_{ij}=E_{ij}+f_{ij}$. Finally, substituting $A_\pm$ back in the master action, produces the T-dual action
\bea
S_{\rm dual}&=&\int d^2\sigma[Q_{\mu\nu}\d_+X^\mu \d_- X^\nu
+(\d_+v_i+Q_{\mu i}\d_+X^\mu)\times \cr
&&\times M^{-1}_{ij}(\d_-v_j-Q_{j\mu}\d_- X^\mu)].
\eea
At the quantum level, i.e. considering the one-loop determinant, the T-duality also modifies the dilaton to
\be
\Phi(x,v)=\Phi(x)-\frac{1}{2}\ln(\det M).
\ee

\section{Particle-Vortex duality as Nonabelian T-duality in 3 dimensions}

We now want to generalize the above construction to $(2+1)-$dimensions. Again, it is natural to consider the {\em real} variables $\Phi_{0}^{k}$ and
$
L_\mu^i=-i\Tr[t^i g^{-1}\d_\mu g]\;,
$
where, as before $g(x^{\mu})\in SU(2)$ is complex. We will first write down a desired master action generalizing the $2-$dimensional case, except 
with $Q_{\mu i}=0$ and $Q_{\mu\nu}=\delta_{\mu\nu}$.  First though, we define the local $SU(2)$ symmetry, which means replacing derivatives 
with covariant derivatives, $D_\mu g=\d_\mu g-A_\mu g$, and $L_\mu^i$ with $\tilde L_\mu^i=-i\Tr[t^i g^{-1}D_\mu g]$. The desired master action is then
\bea
S_{\rm master}&=&\int d^3x \Bigl[-\frac{1}{2}(\d_\mu \Phi_0^k)^2-\frac{1}{2}(\Phi_0^k)^2g^{\mu\nu}\tilde L_\mu^i \tilde L_\nu^j
E_{ij}\cr
&+& \epsilon^{\mu\nu\rho}v_\mu^i F_{\nu\rho}^i\Bigr]\;,\label{master3d}
\eea
where the gauge field strength is the usual $F_{\mu\nu}=\d_\mu A_\nu -\d_\nu A_\mu -[A_\mu,A_\nu]$.

Varying the action with respect to the Lagrange multipliers $v_\mu^i$ leads to $F_{\mu\nu}^i=0$ which, in the absence of any topological issues, 
leads to a trivial gauge field. Consequently, the choice of $A_\mu=0$ leads to $\tilde L_\mu^i=L_\mu^i$, reducing the action to the pre-dualizing,
\be
S_{\rm original}=\int d^3x \left[-\frac{1}{2}(\d_\mu\Phi_0^k)^2-\frac{1}{2}(\Phi_0^k)^2g^{\mu\nu}L_\mu^i L_\nu^j 
E_{ij}\right].\label{original3d}
\ee
If instead we first partially integrate the Lagrange multiplier term to 
\be
\int \epsilon^{\mu\nu\rho} v_\mu^i F_{\nu\rho}^i=
\int \epsilon^{\mu\nu\rho}[(\d_\mu v_\nu^i)A_\rho^i-(\d_\nu v_\mu^i)A_\rho ^i
+A_\mu^i f_{\nu ij}A_\rho^j]\;,\nonumber
\ee
where 
$
f_{\nu ij}\equiv f_{ijk}v_\nu^k\;,
$
and gauge fix by setting $g=1$, then 
$
\tilde L_\mu^i\rightarrow i\Tr[t^i A_\mu]=iA_\mu^i\;.
$
Subsequent variation of the master action with respect to $A_\mu^i$ gives
\be
[(\Phi_0^k)^2g^{\mu\rho}E_{ij}+2\epsilon^{\mu\nu\rho}f_{\nu ij}]A_\rho^j=-\epsilon^{\mu\nu\rho}
(\d_\nu v_{\rho i}-\d_\rho v_{\nu i}),\nonumber
\ee
which is solved by 
$
A^{i}_\mu = -{M^{-1}_{ij}}^{\mu\rho}V_j^\rho\;,\label{AMV}
$
with
\bea
M_{ij}^{\mu\rho}&\equiv& [(\Phi_0^k)^2g^{\mu\rho}E_{ij}+2\epsilon^{\mu\nu\rho}f_{\nu ij}]\cr
V_i^\mu&\equiv & \epsilon^{\mu\nu\rho}(\d_\nu v_{\rho i}-\d_\rho v_{\nu i}).\label{MV}
\eea
On substituting $A^{i}_\mu$ back in the master action (\ref{master3d}), we get the particle-vortex dual action
\bea
S_{\rm dual}&=& \int d^3x \left[-\frac{1}{2}(\d_\mu\Phi_0^k)^2+\frac{1}{2}A_\mu^i M_{ij}^{\mu\rho}A_\rho^j+A_\mu^i V_i^\mu\right]\cr
&=&-\frac{1}{2}\int d^3x [V_i^\mu M^{-1}_{ij}V_j^\rho +(\d_\mu \Phi_0^k)^2].
\eea
Evidently then, we have found a transformation of the path integral in $(2+1)-$dimensional theories of the form (\ref{original3d}) that furnishes a 
{\it non-abelian} particle-vortex duality. In order to 
consider it a genuine particle-vortex duality transformation, we must be able to derive (\ref{original3d}) from a more familiar action that admits 
vortex solutions, couple the theory to a nontrivial gauge field and add a vortex current term to the action. 

To show that this sequence can be executed, we consider a scalar field  $\Phi$ in a tensor product representation, obtained from the adjoint 
representations of two groups, 
that {\em a priori} need not be related to the $SU(2)$ on which particle-vortex duality acts. As an ansatz we take 
\be
\Phi=\Phi_0^a \,\,T_a\otimes e^{i\int dx^\mu L_\mu^i F_i^A \tilde T_A}\;,
\ee
where $T_a$ and $\widetilde{T}_A$ are adjoint matrices transforming under {\it a priori} different groups, and $F_i^A$ are given coefficients (a "background"), 
out of which we will construct $E_{ij}$. Normalizing the generators through 
$\Tr[T_a T_b]=\delta_{ab}$ and $\Tr[\widetilde{T}_A\widetilde{T}_B]=\delta_{AB}$, leads to 
\be
\Tr[(T_a\otimes \tilde T_A)(T_b\otimes \tilde T_B)]=\delta_{AB}\delta_{ab}\;,
\ee
and consequently, the standard kinetic term for $\Phi$ becomes ($\delta^A_A\equiv N$)
\be
\Tr|\d_\mu\Phi|^2=N(\d_\mu\Phi_0^a)^2+(\Phi_0^a)^2L_\mu^i L_\mu ^j N\, E_{ij}\;,
\ee
where $N\, E_{ij}\equiv F_i^A F_j^A$, which up to a normalization of $\Phi_0$ is the same as (\ref{original3d}). We can now add to this action 
a potential depending only on $\Phi_0^a$ which, as we saw earlier, is untouched by the duality transformation. Thereafter, we need to couple 
to a gauge field, write a vortex ansatz and add a vortex current to the action. Toward this end, we need a more general ansatz for the scalar. 

One simple, if naive, possibility is if $F_i^A$ is simply $F_i$, i.e. $T_A$ is trivial and in which we can write an ansatz with just a common phase,
\be
\Phi^{a}=\Phi_0^a \exp\left(i \int dx^\mu L_\mu^i F_i\right),\label{phiasimple}
\ee
and for which the standard scalar kinetic term becomes
\be
\sum_{a}|\d_\mu \Phi^{a}|^2=(\d_\mu\Phi_0^a)^2+(\Phi_0^a)^2 L_\mu^i L_\nu^jg^{\mu\nu}F_iF_j\;.
\ee
Again, we reproduce (\ref{original3d}) except with $E_{ij} = F_{i}F_{j}$ now separable. Next, we couple the scalar to an external gauge 
field, $a_\mu=a_\mu^m T_m$ in a Lie algebra direction not covered by $A_\mu$ ($\Tr[A_\mu T_m]=0$). 
This amounts to replacing $\tilde L_\mu^i$ in (\ref{master3d}) by 
\be
\tilde{\tilde L}_\mu^i=-i\Tr[t^i g^{-1}(\d_\mu-i(A_\mu+a_\mu^mT_m)) g]
\ee
and adding a kinetic term of $+\frac{1}{4}\Tr[f_{\mu\nu}^2]$, for the external gauge field. 

However, for the purposes of writing a vortex ansatz, it is more useful to consider instead a modification that creates a covariant 
derivative acting on the field $\Phi$. For $\Phi$ in the adjoint representation, the normal derivative is 
\be
\d_\mu \Phi=(T_a\d_\mu \Phi_0^a +T_a\otimes T_A i\Phi_0^a L_\mu^i F_i^A){\bf 1}\otimes e^{i\int dx^\mu L_\mu^i F_i^A T_A}\;,
\ee
Making the derivative covariant results in 
\bea
D_\mu\Phi&=&(T_a\d_\mu \Phi_0^a +T_a\otimes T_A i\Phi_0^a L_\mu^i F_i^A +
T_a\Phi_0^a\otimes\cr
&& [A_\mu^BT_B,e^{i\int dx^\mu L_\mu^i F_i^A T_A}]e^{-i\int dx^\mu L_\mu^k F_k^A T_A})\cr
&&{\bf 1}\otimes e^{i\int dx^\mu L_\mu^j F_j^A T_A}
\eea
Therefore, in effect, the gauge field coupling gives the replacement
\be
L_\mu ^i F_i^A\rightarrow L_\mu^i F_i^A +L_\mu^i F_i^B {f_{BC}}^A A_\mu ^C+{\cal O}((L_\nu^j)^2)\;,
\ee
to first order. We note that nothing makes it necessary that the gauge field be nonabelian at all. Indeed, if $A$ belongs to the singlet 
representation, we may write the usual $U(1)$ covariant 
derivative for $\Phi$ without a problem.

We are now ready to consider a vortex ansatz. Assuming azimuthal symmetry, $\Phi_0^a=\Phi_0^a(r)$ and ``vorticial" information 
about the solution is encoded in its phase
\be
  e^{i\int dx^\mu L_\mu^i F_i^A} =e^{iN_A\theta}\;,\label{vortex}
\ee
where $N_A$ is the vortex number and $\theta$ is the polar angle on the plane. For a $U(1)$ gauge field, it suffices to simply erase the 
$A$ index. As in the abelian case, the requirement that $D_\mu\Phi\rightarrow 0$ at $r\rightarrow \infty$ ensures both a finite energy 
solution (since the kinetic term $|D_\mu\Phi|^2$ vanishes at infinity) and the existance of a topological charge (since it implies that 
$\oint A_\theta d\theta$ is quantized). Of course, having an ansatz doesn't guarantee the existence of a solution. One needs to show 
that it is a solution of the equations of motion in a specific model (specified by a particular potential $V(\Phi_0^a)$). In a forthcoming 
article, we will show explicitly how the duality acts of nonabelian vortices in an $SU(2)\times U(1)$ gauge theory that arises, for example, 
in the low energy limit of $\mathcal{N}=2$, $SU(3)$ QCD with $N_{f}$ flavors \cite{Auzzi:2003fs}. 

Finally, with an actual solution at hand we can isolate the vortex contributions to the action in the path integral, and obtain a vortex current term. 
Similarly to the abelian case in considered at length in \cite{Murugan:2014sfa}, where the phase $\a$ separates into $ \a_{\rm smooth}+\a_{\rm vortex}$, 
with $\a_{\rm vortex}$ being the part that contains a topological charge of the vortex, we now replace $L_\mu^i$ with 
$L^i_{\mu,{\rm smooth}}+L^i_{\mu, {\rm vortex}}$. Gauge fixing $g=1$, we get $L_\mu ^i =iA_\mu ^i +L^i_{\mu, {\rm vortex}}$, or rather
$
A_\mu^i\rightarrow A^i_{\mu, {\rm smooth}}+A^i_{\mu,{\rm vortex}}.
$
Then, varying the master action (\ref{master3d}) with respect to $A_{\mu, {\rm smooth}}^i$ gives
\be
  A^i_{\mu, {\rm smooth}}+A^i_{\mu,{\rm vortex}}=-{M^{-1}_{ij}}^{\mu\rho}V_j^\rho.
\ee
The associated vortex current term, 
\be
\epsilon^{\mu\nu\rho}v_\mu^i (\d_\nu A^i_{\rho, {\rm vortex}}-\d_\rho A^i_{\nu,{\rm vortex}})\equiv v_\mu^i j^{\mu i}_{\rm vortex}\;,\label{duality}
\ee 
is obtained from the term linear in $A_{\mu}$. From the vortex ansatz (\ref{vortex}), we have 
\be
L_{\mu,{\rm vortex}}^iF_i^A=N^A\d_\mu\theta=N^A\frac{1}{2(\Phi_0^a)^2}j_\mu\;,
\ee
where $j_\mu=\Phi^\dagger \d_\mu\Phi-\Phi\d_\mu\Phi^\dagger $ is a $U(1)$ scalar particle current. 
In other words, the relation 
(\ref{duality}) expresses a duality between particle and vortex currents, 
generalizing the $\epsilon^{\mu\nu\rho}\d_\nu j_\rho=j^\mu_{\rm vortex}$ relation from the abelian case, 
and justifying us calling it a {\it nonabelian particle-vortex duality} for the path integral transformation.\\

\section{An example: semilocal vortices}

To illustrate the above, we now exhibit the duality transformation explicitly for the case of the {\it semilocal (cosmic) strings} 
of \cite{Vachaspati:1991dz, Hindmarsh:1991jq, Gibbons:1992gt}. Defined through the Lagrangian
\begin{eqnarray}
  {\mathcal L} = -\frac{1}{2}|D_{\mu}\Phi|^{2} - \frac{\lambda}{4}\left(\Phi^{\dagger}\Phi - v^{2}\right)^{2} - \frac{1}{4}f_{\mu\nu}f^{\mu\nu},
\end{eqnarray}
the model is a two-flavored Higgs model with an $SU(2)_{G}\times U(1)_{L}\rightarrow U(2)$ symmetry group. Now the scalar $\Phi=(\Phi^a)=\left(\Phi^1, \Phi^{2}\right)^{T}$
transforms in the {\em fundamental} representation of the global, flavor $SU(2)$, while the gauge-covariant derivative is only $U(1)$-local, 
$D_\mu \Phi=(\d_\mu -ie a_\mu)\Phi$, like at the end of the last section, and 
 $f_{\mu\nu} = 2\partial_{[\mu}a_{\nu]}$ is the usual abelian field strength.
Of course, unlike the case in the last section, where $\Phi=\Phi^a T_a$, so $\Phi$ was in the adjoint of the group generated by $T_a$, now we have a 
scalar $\Phi^a$ in the fundamental representation of the global $SU(2)$, so for the duality transformation we simply write the ansatz
(\ref{phiasimple}) but {\em without} $\Phi=\Phi^aT_a$.
Here $\Phi_0^a$, $a=1,2$ and $L_\mu^i$, $i=1,2,3,4\in adj(U(2))$ are real, $i=4$ corresponds to $\one$, 
thus we see that even though we have 6 real variables, we are constrained to 
have the {\em same} phase for $\Phi^1$ and $\Phi^2$. That is actually fine, since for the axially symmetric $n$-vortex ansatz  
\be
a_\theta=\frac{v}{\sqrt{2}}\frac{n}{r}a(r);\;\;\; a_r=0;\;\;\;
\Phi^a= v \varphi^a(r)e^{in\a_a}\;,
\ee
where $(r,\theta)$ are polar coordinates on the plane, leads to the condition that {\em at $r\rightarrow \infty$}, $\a_2=\a_1+c$, with $c$ a constant.
Taking $c=0$ (without loss of generality), the vortex solution indeed
satisfies the ansatz for the particle-duality transformation in (\ref{phiasimple}).
The energy  is Bogomolnyi-saturated at critical coupling 
 $\beta\equiv 2\lambda/e^{2} =1$, where the second order equations of motion for $\Phi$ and $a_{\mu}$, defining 
 $\varphi(r)=\sqrt{(\varphi^1(r))^2+(\varphi^2(r))^2}$ ,  descend to the first order BPS equations
\be
\frac{d\varphi}{dr}=\frac{n}{r}(1-a)\varphi,\;\;\;\;
\frac{da}{dr}=\frac{r}{n}(1-\varphi^2)\;,
\ee
same ones as for the Nielsen-Olesen vortex, thus the same numerical vortex solution is used to constuct this 
"semi-local string".

Making the identification $\Tr[t^i T_m]=\delta^i_m$ and the embedding $a_\mu^4=a_\mu$, $a_\mu^{1,2,3}=0$ (and $A_\mu^{1,2,3}\neq 0$; $A_\mu^3=0$), 
we have the master
action for the duality (replacing $\tilde L_\mu^i$ with $\tilde{\tilde L}_\mu^i$ in (\ref{master3d}) and adding the kinetic term)
\bea
S_{\rm master}&=&\int d^3x \Bigl[-\frac{1}{2}(\d_\mu \Phi_0^a)^2-\frac{1}{2}(\Phi_0^a)^2g^{\mu\nu}\sum_{i,j=1}^4\tilde{\tilde L}_\mu^i \tilde{\tilde L}_\nu^j
E_{ij}\cr
&-& \frac{1}{4}f_{\mu\nu}^2-V(\Phi)+\epsilon^{\mu\nu\rho}\sum_{i=1,2,3}v_\mu^i F_{\nu\rho}^i\Bigr]\;,
\eea
where $E_{ij}=F_iF_j$ and $\Phi_{0}^{a} = v\varphi^{a}$. As before, varying with respect to $v_\mu^i$ 
leads to the original action, where the terms on the first line combine to 
give $-(1/2)|D_\mu\Phi|^2$. Integrating out $A_\mu$ instead and imposing the gauge $g=1$, leads to the dual action  (with the definitions (\ref{MV}))
\bea
&&S_{\rm dual}=\int d^3x\left[-\frac{1}{2}(\d_\mu \Phi_0^a)^2-\frac{1}{4}f_{\mu\nu}^2-V(\Phi)+A_\mu^i V_i^\mu\right.\cr
&&\left.+A_\mu^{\tilde i}(V_{\tilde j}^\rho+M_{\tilde i 4}^{\rho\sigma}a_\sigma)+\frac{1}{2}a_\mu g^{\mu\rho}(\Phi_0^a)^2 a_\rho
+\frac{1}{2}A_\mu^{\tilde i} M_{\tilde i\tilde j}^{\mu\rho}A_\rho ^{\tilde j}\right]\;,\cr\nonumber
\eea
where $A_\mu^{\tilde i}=-M_{\tilde i\tilde j}^{-1\mu\rho}(V_{\tilde j}^\rho+M_{\tilde i 4}^{\rho\sigma}a_\sigma)$.

\section{Discussion}

Abelian particle-vortex duality has proven a powerful tool in the understanding of bosonic systems that range from anyonic superconductivity 
through to cosmic strings. An excellent example of this is illustrated in \cite{Burgess:2001sy}, which utilizes precisely this duality to explain 
the current-voltage symmetry observed near the critical point of the transition between the Laughlin plateaux and Quantum Hall insulator, 
a phenomenon not captured in the linear electromagnetic approximation. 

As exciting as these developments have been to date, we are today at the birth of a new scientific paradigm with the discovery of topological 
phases of matter as embodied in, for example, high temperature superconductors and the fractional quantum Hall effect. A key feature of such 
states of matter is that their quasi-particle excitations are neither fermionic nor bosonic but are best described as {\it nonabelian anyons} 
that obey nonabelian braiding statistics. Certainly since Moore and Read's landmark paper \cite{Moore:1991ks} identifying quasiparticle 
excitations of certain fractional quantum Hall systems which obey nonabelian statistics, nonabelian states of matter have posed an exciting 
challenge to theoretical physics. Recent technological advances coupled with equally rapid developments in topological field theory have 
served only to fuel interest in this area and make the study of nonabelian states of matter one of the hottest topics in theoretical condensed 
matter physics today. It is our hope that the nonabelian particle-vortex duality communicated 
in this article will develop into as useful a tool to understand these new states of matter as its counterpart did for abelian physics.

\section{Acknowledgements}
We thank Thiago Ara\'{u}jo, Fernando Quevedo and Jonathan Shock for discussions.
The work of HN is supported in part by CNPq grant 301709/2013-0 and FAPESP grants 2013/14152-7 and 2014/18634-9.
JM acknowledges support from  the National Research Foundation (NRF) of South Africa under its IPRR and CPRR programs.

\end{document}